\documentclass[conference]{IEEEtran}
\IEEEoverridecommandlockouts
\usepackage[T1]{fontenc}
\usepackage{multirow}
\usepackage{makecell}
\usepackage{multirow}
\usepackage{amsmath,amssymb,amsfonts}
\usepackage{algorithmic}
\usepackage{algorithm}
\usepackage{array}
\usepackage{orcidlink}
\usepackage[caption=false,font=normalsize,labelfont=sf,textfont=sf]{subfig}
\usepackage{textcomp}
\usepackage{stfloats}
 \usepackage{booktabs}
\usepackage{verbatim}
\usepackage{graphicx}
\usepackage{amsmath}
\usepackage{amsfonts}
\usepackage{graphicx}
\usepackage{tikz}
\usepackage{hyperref}
\usepackage[table]{xcolor}
\usepackage[printonlyused, nolist]{acronym}
\usepackage{float} 
\usepackage{tikz}
\hypersetup{hidelinks}

\newcommand\submittedtext{%
  \footnotesize This work has been submitted to the IEEE for possible publication. Copyright may be transferred without notice, after which this version may no longer be accessible.}

\newcommand\submittednotice{%
\begin{tikzpicture}[remember picture,overlay]
\node[anchor=south,yshift=10pt] at (current page.south) {\fbox{\parbox{\dimexpr0.65\textwidth-\fboxsep-\fboxrule\relax}{\submittedtext}}};
\end{tikzpicture}%
}
\def\BibTeX{{\rm B\kern-.05em{\sc i\kern-.025em b}\kern-.08em
    T\kern-.1667em\lower.7ex\hbox{E}\kern-.125emX}}

\bstctlcite{IEEEexample:BSTcontrol}
\begin{document}
\title{LITE: Lightweight Channel Gain Estimation with Reduced X‑Haul CSI Signaling in O‑RAN}

 \author{
    \IEEEauthorblockN{
        David Góez\orcidlink{0000-0001-7658-0994}\IEEEauthorrefmark{1},
        Marco Piazzola\IEEEauthorrefmark{2},
        Giulia Costa\orcidlink{0000-0002-0815-9652}\IEEEauthorrefmark{2}
        Achiel Colpaert\orcidlink{0000-0003-1238-2002}\IEEEauthorrefmark{3}
        Rodney Martinez Alonso\orcidlink{0000-0003-2529-5944}\IEEEauthorrefmark{4}\\
        Esra Aycan Beyazıt\orcidlink{0000-0003-1035-6695}\IEEEauthorrefmark{1}  
        Nina Slamnik-Kriještorac\orcidlink{0000-0003-1719-772X}\IEEEauthorrefmark{1},
        Johann M. Marquez-Barja\orcidlink{0000-0001-5660-3597}\IEEEauthorrefmark{1},
        Miguel Camelo Botero\orcidlink{0000-0001-8152-7143}\IEEEauthorrefmark{1}}\\
    \IEEEauthorblockA{\IEEEauthorrefmark{1}University of Antwerp - imec, IDLab, Belgium}
    \IEEEauthorblockA{\IEEEauthorrefmark{2}Spindox Labs SRL, Italy}
    \IEEEauthorblockA{\IEEEauthorrefmark{3}imec, Kapeldreef 75, 3001 Leuven, Belgium}
    \IEEEauthorblockA{\IEEEauthorrefmark{4}ESAT, KULEUVEN, Leuven, Belgium}
    \vspace*{-2em}
  \thanks{This work is supported by the 6G-BRICKS project, which has received funding from the European Union’s Horizon Europe program under Grant Agreement No 101096954 and by the Flemish Government under the “Onderzoeksprogramma Artificiële Intelligentie (AI) Vlaanderen” program. The work of Rodney Martinez Alonso is supported by the Research Foundation–Flanders (FWO) under Grant 1211926N.}
  }

\maketitle
\submittednotice

\begin{abstract}
\ac{CF-MaMIMO} in \ac{O-RAN} promises high spectral efficiency but is limited by frequent \ac{CSI} exchanges, which strain \ac{X-haul} bandwidth and exceed the capabilities of existing approaches relying on uncompressed \ac{CSI} or heavy predictors. To overcome these constraints, we propose \acs{LITE}, a lightweight pipeline combining a 1-D convolutional \ac{AE} at the \ac{O-DU} with a \ac{SE}-enhanced \ac{BiLSTM} predictor at the \ac{Near-RT-RIC}, enabling short-horizon trajectory-unaware forecasting under strict transport and processing budgets. LITE applies $50\%$ \ac{CSI} compression and an asymmetric \ac{SE}-BiLSTM, reducing model complexity by $83.39\%$ while improving accuracy by $5\%$ relative to a baseline BiLSTM. With compression-aware training, the \ac{LITE} incurs only $6\%$ accuracy loss versus the BiLSTM baseline, outperforming independent and end-to-end strategies. A TensorRT-optimized implementation achieves $147k$~\ac{QPS}, a 4.6x throughput gain. These results demonstrate that LITE delivers \ac{X-haul}-efficient, low-latency, and deployment-ready channel-gain prediction compatible with O-RAN splits.
\end{abstract}

\begin{IEEEkeywords}
Channel Estimation, 5G-NR, OFDM, Deep Learning, Neural Attention, Neural Network Acceleration.
\end{IEEEkeywords}

\begin{acronym}
    \acro{LITE}{Lightweight Intelligent Trajectory Estimator}
    \acro{CF-MaMIMO}{Cell-Free Massive Multiple-Input Multiple-Output}
    \acro{MIMO}{Multiple-Input Multiple-Output}
    \acro{RAN}{Radio Access Network}
    \acro{O-RAN}{Open Radio Access Network}
    \acro{AP}{Access Point}
    \acro{UE}{User Equipment}
    \acro{CSI}{Channel State Information}
    \acro{O-DU}{O-RAN Distributed Unit}
    \acro{O-RU}{O-RAN Radio Unit}
    \acro{RIC}{RAN Intelligent Controller}
    \acro{Near-RT-RIC}{Near-Real-Time RAN Intelligent Controller}
    \acro{xApp}{RIC-native Application}
    \acro{E2}{E2 Interface}
    \acro{O1}{O1 Management Interface}
    \acro{O2}{O2 Orchestration Interface}
    \acro{O3}{O3 Interface}
    \acro{O4}{O4 Interface}
    \acro{OFDM}{Orthogonal Frequency-Division Multiplexing}
    \acro{SNR}{Signal-to-Noise Ratio}
    \acro{AE}{Autoencoder}
    \acro{DAE}{Deep Autoencoder}
    \acro{RRM}{Radio Resource Management}
    \acro{KPI}{Key Performance Indicator}
    \acro{KPM}{Key Performance Measurement}
    \acro{RC}{RAN Control (Service Model)}
    \acro{Midhaul}{Transport segment between O-DU and Near-RT RIC}
    \acro{LITE-Compress}{LITE Compression Module}
    \acro{LITE-Decompress}{LITE Decompression Module}
    \acro{DNN}{Deep Neural Network}
    \acro{X-haul}{fronthaul/midhaul/backhaul}
    \acro{SE}{Squeeze-and-Excitation}
    \acro{MSE}{Mean Squared Error}
    \acro{RMSE}{Root Mean Squared Error}
    \acro{BiLSTM}{Bidirectional Long Short-Term Memory}
    \acro{E2E}{End-to-End}
    \acro{FC}{Fully-Connected/Dense}
    \acro{GPU}{Graphics Processing Unit}
    \acro{QPS}{Queries per Second}
    \acro{ML}{Machine Learning}
    \acro{DL}{Deep Learning}
\end{acronym}
\section{Introduction}\label{sec:intro}
\IEEEPARstart{A}{s} wireless connectivity demand continues to surge, next-generation networks must deliver high spectral efficiency, robust mobility support, and uniform quality of experience \cite{ericsson2025mobilityreport}. \ac{CF-MaMIMO} has emerged as a promising architecture by combining massive \ac{MIMO} array gains with distributed coverage \cite{Mohammadi2024}, jointly serving all users across shared time–frequency resources. By removing cell borders, \ac{CF-MaMIMO} mitigates inter-cell interference and improves rate fairness, with reported spectral-efficiency gains up to 95\% \cite{Liu2020}.

Despite its potential, scaling \ac{CF-MaMIMO} presents practical challenges. Computational and coordination overheads grow with the number of \acp{AP} and \acp{UE}, while frequent transport of high-dimensional \ac{CSI} over the \ac{X-haul} can exceed realistic bandwidth limits \cite{Björnson2020}. User-centric clustering and scheduling strategies have been proposed to mitigate these constraints \cite{Chen2022,Björnson2020,Chen2021}, yet they often rely on quasi-static assumptions and fail to fully account for dynamic mobility, causing performance degradation when trajectories are unknown \cite{Jiang2022}. Recent works have explored short-term channel-gain prediction from CSI evolution rather than explicit spatial coordinates \cite{Alonso2024}, but these methods either use uncompressed \ac{CSI}, imposing excessive transport load, or rely on computationally heavy predictors unsuitable for real-time O-RAN deployment.

The \ac{O-RAN} architecture offers a flexible and open framework for deploying data-driven control in \ac{CF-MaMIMO}, leveraging disaggregated processing and standardized interfaces \cite{Beerten2025}. However, its distributed nature intensifies \ac{X-haul} limitations: frequent \ac{CSI} transport or measurement reporting can overload midhaul links. Prior studies investigated functional split optimization \cite{Girycki2024} and lightweight \ac{CSI} compression \cite{Liu2022,Mismar2024}, yet the combined impact of compression and predictive modeling on short-horizon channel-gain forecasting remains insufficiently characterized.


To address these gaps, we introduce \ac{LITE}, an end‑to‑end pipeline for trajectory‑unaware channel‑gain prediction in \ac{CF-MaMIMO} under \ac{O-RAN} constraints. \ac{LITE} integrates a compact 1‑D convolutional \ac{AE} at the \ac{O-DU} for \ac{CSI} compression with an asymmetric, \ac{SE}-enhanced \ac{BiLSTM} predictor at the \ac{Near-RT-RIC}. This architecture is designed to lower transport overhead and computational complexity while enabling accurate short‑horizon channel prediction without requiring explicit trajectory information. Compared with existing symmetric \ac{BiLSTM}-based approaches, \ac{LITE} provides a more efficient model structure and supports real‑time inference within \ac{O-RAN} processing constraints.

The remainder of the paper is structured as follows. Section~\ref{sec:architecture} details the LITE system architecture. Section~\ref{sec:algorithms} presents the CSI compression and prediction algorithms. Section~\ref{sec:results} reports experimental results, and Section~\ref{sec:conclusion} concludes the paper and outlines directions for future work.

\section{LITE Architecture}\label{sec:architecture}
\begin{figure*}[!t]
    \centering
    \includegraphics[width=0.8\textwidth]{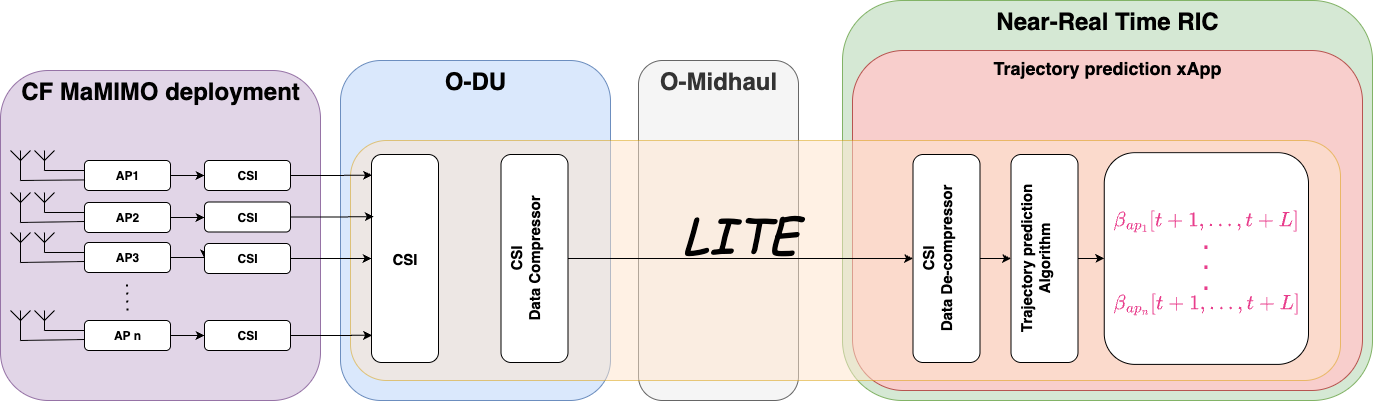}
    \caption{Overview of the LITE system architecture, illustrating the four processing layers}
    \label{fig:LITE_arch}
\end{figure*}

As a first step toward realizing \ac{LITE}, a compact system architecture is designed and illustrated in Fig.~\ref{fig:LITE_arch}. The framework operates across a \ac{CF-MaMIMO} deployment and an \ac{O-RAN} \ac{Near-RT-RIC} environment while maintaining full compatibility with existing \ac{O-RAN} interfaces. No modifications are introduced to the \acp{AP} or \acp{O-RU}, ensuring seamless integration with standard RAN processing pipelines.

The architecture comprises the fololwing four tightly coupled processing layers:

\subsubsection*{\textbf{\ac{CF-MaMIMO} Radio Access Layer}}
Uplink signals transmitted by user devices are received by a distributed array of \acp{AP}, from which standard pilot-based estimation yields full complex-valued \ac{CSI} following the \ac{O-RAN}~7.2x functional split. As shown in \cite{Alonso2024}, the resulting tensor of the channel can be represented as:
\begin{equation}
H(t) \in \mathbb{C}^{N_{\mathrm{AP}}\times N_{\mathrm{ant}}\times N_{\mathrm{sb}}},
\end{equation}
For a flat fading channel, this can be further reduced to a per-\ac{AP} large-scale channel-gain vector via antenna–subcarrier averaging:
\begin{equation}
\beta(t) \in \mathbb{R}^{N_{\mathrm{AP}}}.
\end{equation}

To ensure a uniform temporal structure, the sequence $\boldsymbol{\beta}(t)$ is reorganized into a fixed-size window representation:
\begin{equation}
S \in \mathbb{R}^{N_{\mathrm{AP}}\times Z},
\end{equation}
where $Z$ denotes the temporal window length, corresponding to the number of consecutive CSI snapshots
$\{\boldsymbol{\beta}(t-Z+1), \ldots, \boldsymbol{\beta}(t)\}$ stacked along the time dimension.
This representation serves as the canonical input to the \ac{LITE} processing chain, while the raw tensors $H(t)$ remain accessible for validation and reference.

\subsubsection*{\textbf{\ac{O-DU} Processing Layer}}
At the \ac{O-DU}, the harmonized representation \(S\) is normalized and passed to a \ac{DNN}-based \ac{AE} that learns a compact latent encoding. This generates a low-dimensional representation \(L(t)\) satisfying \(|L| \ll |H|\), enabling efficient transport while preserving the temporal and spatial characteristics necessary for prediction.

\subsubsection*{\textbf{Midhaul Transport Layer}}
The latent representation is forwarded to the \ac{Near-RT-RIC} via the \ac{E2} interface. The sharp reduction in dimensionality alleviates midhaul bandwidth consumption and aligns with the \ac{O-RAN} disaggregated processing paradigm.

\subsubsection*{\textbf{\ac{Near-RT-RIC} Execution Layer}}
Within the \ac{RIC}, the latent representation is decoded to reconstruct a high-resolution sequence suitable for temporal analysis. A prediction module then processes this reconstructed sequence to forecast short-horizon channel dynamics relevant for downstream \ac{RRM} functions such as scheduling or beamforming. The \ac{xApp} includes:
\begin{itemize}
  \item \ac{CSI} Decompression (LITE-Decompress)
  \item Temporal Dynamics Modeling
  \item Short-Horizon \ac{CSI} Prediction
  \item \ac{RIC} Output Adaptation
\end{itemize}

The processing chain thus maps raw \ac{CSI} measurements $H(t)$, locally estimated at the radio access layer, to future channel-state predictions consumed by \ac{RIC}-hosted control functions. The harmonized temporal sequences \(S\) form the stable input domain, and the predicted channel states form the actionable output.

Realizing this \ac{E2E} functionality requires jointly optimized learning modules that operate under strict dimensionality and accuracy constraints. A \ac{DNN}-based encoder produces a compact representation for efficient transport, a decoder reconstructs a high-resolution feature space required for reliable temporal modeling, and a predictor processes this reconstructed sequence using architectures trained on the full dataset. This design ensures that compression reduces only transport overhead, without compromising predictive capability. The encoder, decoder, and predictor must therefore satisfy stringent \acp{KPI}, including high compression efficiency, high reconstruction fidelity, and robust prediction accuracy under aggressive compression. Together, these components constitute the core of the \ac{LITE} \ac{E2E} learning model.
\section{CSI compression and channel gain prediction}\label{sec:algorithms}
 
\begin{figure}[!t]
    \centering
    \includegraphics[width=\columnwidth]{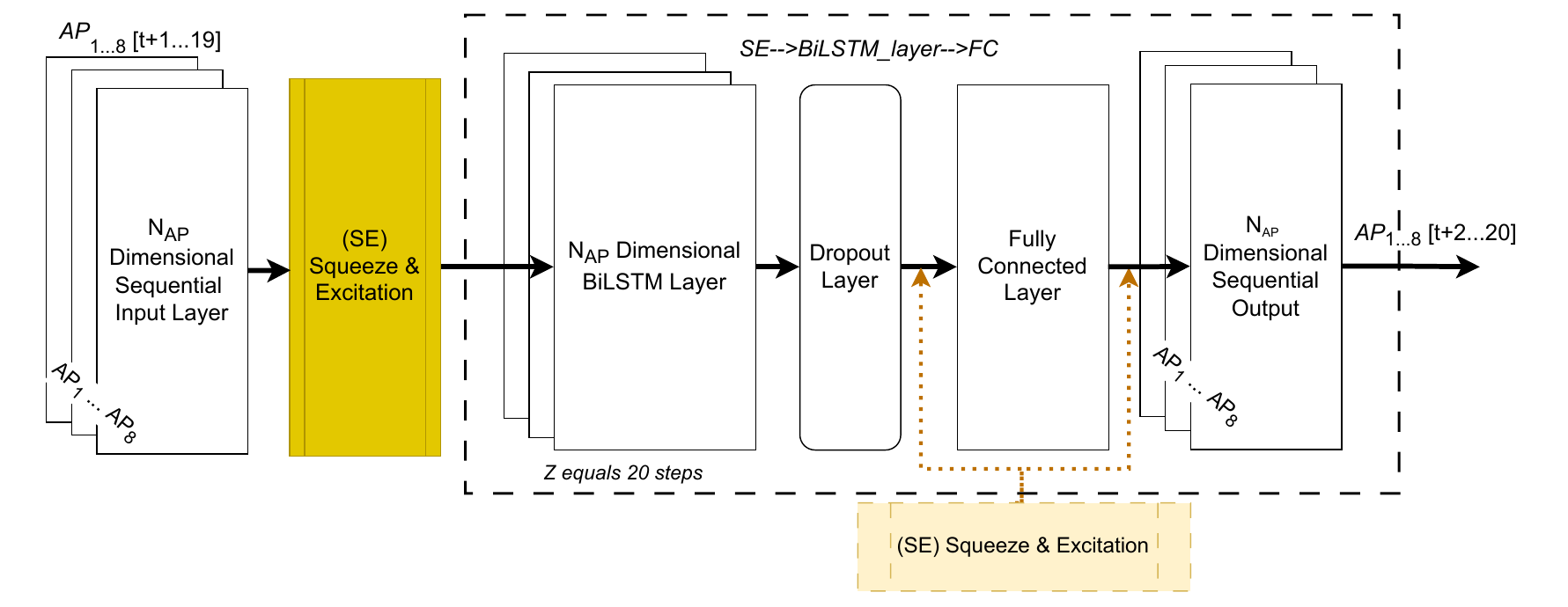}
    \caption{Detailed LITE \ac{DL} architecture showing the symmetric 1‑D convolutional \ac{AE} and the \ac{SE}‑enhanced \ac{BiLSTM} predictor. Optional \ac{SE} blocks are illustrated at intermediate and output stages for comparison}
    \label{fig:LITE_DNN}
\end{figure}

Fig.~\ref{fig:LITE_DNN} introduces the \ac{E2E} \ac{DL} architecture empowering \ac{LITE} that jointly optimizes \emph{CSI compression} and \emph{trajectory-unaware channel-gain forecasting} under strict \ac{Near-RT-RIC} constraints. Unlike prior work, our design combines (i) a convolutional \ac{AE} tailored for temporal CSI sequences and (ii) a lightweight BiLSTM predictor augmented with \ac{SE} attention, achieving substantial compression while preserving predictive accuracy. 

\subsection{Autoencoder for CSI Compression}
\label{subsec:ae}

\ac{X-haul} transport of raw \ac{CSI} tensors $\mathbf{H}(t)\in\mathbb{C}^{N_{\text{AP}}\times N_{\text{ant}}\times N_{\text{sc}}}$ constitutes a major scalability bottleneck. To mitigate this, \textsc{LITE} learns a compact latent representation of per-AP channel-gain sequences using a \ac{DNN}-based \ac{AE}, as such architectures have demonstrated strong performance in radio signal processing tasks~\cite{Camelo2019,camelo2020ai,Fontaine2020}. Specifically, the \ac{AE} maps aggregated \ac{CSI} windows:
\begin{equation}
\mathbf{S}\in\mathbb{R}^{N_{\text{AP}}\times W}
\end{equation}
to a lower-dimensional latent representation:
\begin{equation}
\mathbf{L}(t)\in\mathbb{R}^{N_{\text{AP}}\times Z}, \quad Z \ll W,
\end{equation}
where $\mathbf{S}$ aggregates per-AP large-scale channel gains over fixed-length temporal windows.

The proposed \ac{AE} adopts a lightweight symmetric 1-D convolutional design. As shown in Table~\ref{tab:ae}, the encoder consists of five strided Conv1d layers that progressively reduce the temporal dimension while increasing feature depth, mapping an input of shape $[B,1,152]$ to a latent representation of size $[B,15,5]$ through successive stride-2 temporal downsampling. ReLU activations are used in all intermediate layers, while the latent layer is linear. The decoder mirrors the encoder using ConvTranspose1d layers to restore the original temporal resolution, reconstructing $\hat{\mathbf{S}}\in\mathbb{R}^{[B,1,152]}$.

The \ac{AE} design achieves an approximate $2\times$ reduction in representation size, i.e., a 50\% compression ratio, by compressing the input from 152 to 75 real-valued features. This operating point provides a practical trade-off between \ac{X-haul} bandwidth reduction and reconstruction fidelity, as supported by prior work~\cite{Mismar2024}. It halves the \ac{CSI} payload exchanged between the \ac{O-DU} and the \ac{Near-RT-RIC} while preserving dominant temporal correlations and large-scale fading characteristics relevant for downstream prediction. Fixing the compression ratio further enables a controlled analysis of dataset augmentation and trajectory diversity, without introducing additional architectural degrees of freedom, as will be shown in Section~\ref{sec:results}.

\begin{table}[!t]
\caption{Architecture of the symmetric 1‑D convolutional \ac{AE} used for \ac{CSI} compression, including encoder and decoder layer configurations, kernel sizes, strides, and activation functions}
\label{tab:ae}
\resizebox{\columnwidth}{!}{%
\begin{tabular}{|c|c|c|c|c|c|}
\hline
\textbf{Stage} & \textbf{Type} & \textbf{Channels} & \textbf{Kernel} & \textbf{Stride} & \textbf{Activation} \\ \hline
E1 & Conv1d & $1 \rightarrow 64$    & 5 & 2 & ReLU \\ \hline
E2 & Conv1d & $64 \rightarrow 512$  & 3 & 2 & ReLU \\ \hline
E3 & Conv1d & $512 \rightarrow 256$ & 3 & 2 & ReLU \\ \hline
E4 & Conv1d & $256 \rightarrow 128$ & 3 & 2 & ReLU \\ \hline
E5 & Conv1d & $128 \rightarrow 15$  & 3 & 2 & Linear \\ \hline
D1 & ConvT1d & $15 \rightarrow 128$ & 3 & 2 & ReLU \\ \hline
D2 & ConvT1d & $128 \rightarrow 256$ & 3 & 2 & ReLU \\ \hline
D3 & ConvT1d & $256 \rightarrow 512$ & 3 & 2 & ReLU \\ \hline
D4 & ConvT1d & $512 \rightarrow 64$  & 3 & 2 & ReLU \\ \hline
D5 & ConvT1d & $64 \rightarrow 1$    & 5 & 2 & Linear \\ \hline
\end{tabular}
}
\end{table}

\subsection{Lightweight Attention-Based Predictor}
\label{subsec:predictor}

Trajectory-unaware forecasting must capture non-linear temporal dependencies and inter-AP coupling without relying on explicit position information. Although Transformer variants were considered, their quadratic sequence complexity and memory footprint are ill-suited to \ac{Near-RT-RIC} constraints. \textsc{LITE} therefore adopts a \ac{BiLSTM} \cite{Schuster1997} backbone, which has shown strong accuracy–efficiency trade-offs for \ac{CF-MaMIMO} channel prediction~\cite{Alonso2024}. Bidirectional processing leverages past and future context to limit error accumulation in single-step, short-horizon prediction, and naturally supports multi-AP joint modelling.

To reduce computational load, we employ lightweight and asymmetric \ac{BiLSTM} configurations that decrease the hidden-state dimensionality of the forward/backward paths while maintaining stable accuracy. To further strengthen feature selectivity, a \ac{SE} block~\cite{Hu2018} is inserted \emph{before} the recurrent layers, performing channel-wise reweighting of AP features. This early recalibration dampens noisy or redundant inputs and enables smaller recurrent layers without compromising temporal modelling. Alternative placements were explored, such as after the \ac{BiLSTM} and the regression head, and found to be less effective at preserving temporal dependencies. The adopted placement best balances robustness and efficiency. This architectural decision is validated in Section~\ref{sec:results}.

\subsection{End-to-End Integration and Training}
\label{subsec:e2e}

The \textsc{LITE} pipeline is implemented as a modular yet tightly coupled workflow that spans the O-DU and Near-RT RIC, integrating four stages: \emph{compression $\rightarrow$ midhaul transport $\rightarrow$ decompression $\rightarrow$ prediction}. This design ensures that CSI is compressed at the edge before transport, reconstructed at the RIC, and then consumed by the predictor without introducing distribution mismatches. Achieving this requires careful alignment between the latent representation learned by the autoencoder and the temporal dependencies modeled by the predictor.

To address this, we explored three complementary training strategies:

\begin{enumerate}
    \item \textbf{Independent Training:} The autoencoder and predictor are trained separately on their respective objectives. This approach simplifies optimization and stabilizes convergence, but risks a domain gap because the predictor learns from raw sequences while inference uses reconstructed ones.
    \item \textbf{Compression-Aware Training:} Here, the autoencoder is trained first and frozen, and the predictor is trained on decompressed sequences. This strategy adapts the predictor to compression artifacts without altering the encoder–decoder weights, striking a balance between modularity and robustness. It proved most effective for maintaining prediction accuracy under aggressive compression.
    \item \textbf{End-to-End (Joint) Training:} Both modules are pipelined and trained simultaneously. While conceptually appealing for global optimization, this approach exhibited instability due to conflicting gradients. The reconstruction objective favors smooth latent codes, whereas the forecasting task benefits from preserving fine-grained temporal variations. This tension led to suboptimal latent representations and degraded prediction accuracy, as will be shown in Section~\ref{sec:results}.
\end{enumerate}

From a system perspective, compression-aware training was adopted for deployment because it offers predictable behavior, modular retraining capability, and resilience to X-haul constraints. This design also aligns with O-RAN principles: the encoder runs at the \ac{O-DU} to minimize transport overhead, while the decoder and \ac{SE}-\ac{BiLSTM} predictor can be executed as a containerized xApp within the \ac{Near-RT-RIC}, ensuring portability and providing higher computing capacity compared to the \ac{O-DU}. Together, these choices enable \textsc{LITE} to deliver X-haul-efficient, trajectory-unaware forecasting without compromising integration stability or real-time performance.
\section{Performance Evaluation Results}\label{sec:results}
In this section, we present the performance evaluation of the LITE framework, covering its individual components (\ac{AE} and \ac{SE}-\ac{BiLSTM}) as well as the end-to-end integration.

For \ac{CSI} data, we use the Ultra Dense Indoor MaMIMO CSI Dataset introduced in \cite{DeBast2021}, following the methodology in \cite{Alonso2024}. Since the original traces correspond to static measurements, we apply a data-augmentation procedure based on the virtualized channel gain evolution algorithm from~\cite{Alonso2024}, where the mean variation in channel gain due to user movement, $\Delta \beta$, is modeled as a stochastic process evolving as the user moves from $(X_A, Y_A)$ to $(X_B, Y_B)$ over a time interval $\Delta t$, capturing the spatial dependence of channel variations.

As mentioned in Section~\ref{subsec:ae}, the \ac{AE} uses a fixed 50\% compression ratio, halving the \ac{X-haul} payload while preserving key temporal correlations and large-scale fading characteristics as demonstrated in~\cite{Mismar2024}. Fixing the compression ratio enables a controlled evaluation of reconstruction fidelity, predictor performance, and the impact of dataset augmentation in LITE.

Performance evaluations (Sections~\ref{sec:res:sebilstm}, \ref{sec:res:e2e}, and \ref{sec:res:latency}) use 2500 synthetically generated trajectories, larger than the 200 samples in~\cite{Alonso2024}. The rationale for this dataset size and its influence on the \ac{AE} design are discussed in Section~\ref{sec:res:csi-compre}. The dataset is split into training and validation sets using a 9:1 ratio, as in \cite{Alonso2024}. All models were trained for at least 1000 iterations with early stopping, using a learning rate of 0.01 and a minibatch size of 32, while inference with TensorRT was performed using a batch size of 250 samples.

All experiments were conducted in a Docker container running Ubuntu 20.04.5 LTS, leveraging an NVIDIA GeForce GTX 1650 GPU ($\sim$4,GB VRAM) with CUDA Toolkit 11.8, CUDA Runtime (PyTorch) 11.6, and cuDNN 8.3 for hardware acceleration. The software stack included PyTorch 1.13.1, TensorRT 8.5.1, and ONNX 1.17.0.

\subsection{Channel gain prediction with SE-enhanced BiLSTM}\label{sec:res:sebilstm}
In \textsc{LITE}, we evaluate three \ac{SE} placements, before the \ac{BiLSTM}, after the \ac{BiLSTM} (pre-\ac{FC}), and after the \ac{FC} layer, across multiple asymmetric and symmetric $(f,b)$ hidden-size configurations. The number of forward ($f$) and backward ($b$) hidden units significantly impacts both prediction accuracy and model complexity, as increasing hidden units generally improves temporal modeling, but larger configurations offer diminishing returns relative to parameter growth, especially when \ac{SE} is applied late in the network.

As shown in Table \ref{tab:rmse_complexity_lite}, placing \ac{SE} before the \ac{BiLSTM} consistently provides the most favorable accuracy, complexity trade-off. The asymmetric $(64,128)$ configuration achieves the lowest \ac{RMSE} of \textbf{0.127}, a \textbf{5.03\%} improvement over the baseline, with only $91169$ parameters. A smaller configuration, $(64,96)$, reaches $\mathrm{RMSE}=0.129$ (\textbf{3.57\%} improvement) with just $60961$ parameters, highlighting that moderate backward units effectively enhance temporal encoding without excessive complexity. These results indicate that early channel-wise recalibration enables the \ac{BiLSTM} to focus its limited recurrent capacity on the most informative input dimensions, maximizing the impact of temporal modeling.

\begin{table*}[t]
\centering
\caption{Impact of \ac{SE} block placement on RMSE and model complexity across various BiLSTM hidden‑size configurations. \\ The baseline symmetric (256, 256) model from \cite{Alonso2024} achieves an RMSE of $0.134$.}
\label{tab:rmse_complexity_lite}
\scriptsize
\setlength{\tabcolsep}{3.2pt}
\resizebox{\textwidth}{!}{%
\begin{tabular}{l
                r r r r
                r r r r
                r r r r}
\toprule
& \multicolumn{4}{c}{\textbf{SE Before BiLSTM}}
& \multicolumn{4}{c}{\textbf{SE After BiLSTM (pre-FC)}}
& \multicolumn{4}{c}{\textbf{SE After FC}} \\
\cmidrule(lr){2-5} \cmidrule(lr){6-9} \cmidrule(lr){10-13}
\textbf{LSTM (f,b)}
& \textbf{RMSE} & \textbf{$\Delta$ (\%)} & \textbf{Params} & \textbf{Red. (\%)}
& \textbf{RMSE} & \textbf{$\Delta$ (\%)} & \textbf{Params} & \textbf{Red. (\%)}
& \textbf{RMSE} & \textbf{$\Delta$ (\%)} & \textbf{Params} & \textbf{Red. (\%)} \\
\midrule
(32,32)   & 0.139 & -3.80  & 11297   & 97.94 & 0.144 & -7.52  & 12368   & 97.75 & 0.146 & -9.03 & 11297   & 97.94 \\
(32,64)   & 0.130 & 2.43   & 25121   & 95.42 & 0.134 & 0.54   & 27508   & 94.99 & 0.139 & -3.62 & 25121   & 95.42 \\
(64,32)   & 0.142 & -6.11  & 25121   & 95.42 & 0.146 & -9.13  & 27508   & 94.99 & 0.150 & -12.1 & 25121   & 95.42 \\
(64,64)   & 0.132 & 1.29   & 38945   & 92.90 & 0.138 & -3.26  & 43160   & 92.14 & 0.142 & -5.85 & 38945   & 92.90 \\

(64,96)   & \cellcolor{yellow!20}0.129 & \cellcolor{yellow!20}3.57  
          & \cellcolor{yellow!20}60961 & \cellcolor{yellow!20}88.89
          & 0.132 & 1.11   & 67516  & 87.70 
          & 0.134 & -0.10  & 60961  & 88.89 \\

(96,64)   & 0.134 & -0.20  & 60961   & 88.89 & 0.139 & -4.12  & 67516   & 87.70 & 0.140 & -5.04 & 60961   & 88.89 \\

(64,128)  & \cellcolor{green!20}\textbf{0.127} & \cellcolor{green!20}\textbf{5.03}  
          & \cellcolor{green!20}\textbf{91169} & \cellcolor{green!20}\textbf{83.39}
          & 0.132 & 0.98   & 100576 & 81.68 
          & \textbf{0.133} & \textbf{0.58}  & \textbf{91169}  & \textbf{83.39} \\

(128,64)  & 0.136 & -1.94  & 91169   & 83.39 & 0.141 & -5.11  & 100576 & 81.68 & 0.144 & -7.47 & 91169   & 83.39 \\
(96,128)  & 0.132 & 1.55   & 113185  & 79.38 & 0.134 & -0.17  & 125956 & 77.05 & 0.135 & -0.62 & 113185  & 79.38 \\
(128,96)  & 0.133 & 0.51   & 113185  & 79.38 & 0.137 & -2.17  & 125956 & 77.05 & 0.139 & -4.34 & 113185  & 79.38 \\
(128,128) & 0.131 & 1.70   & 143393  & 73.87 & 0.134 & -0.01  & 160040 & 70.84 & 0.136 & 1.90  & 143393  & 73.87 \\

(128,160) & 0.132 & 1.35   & 181793  & 66.88 
          & \textbf{0.130} & \textbf{2.67}  & \textbf{202828} & \textbf{63.05}
          & 0.136 & -2.05  & 181793  & 66.88 \\

(160,128) & 0.131 & 1.61   & 181793  & 66.88 & 0.134 & -0.39  & 202828 & 63.05 & 0.139 & -3.67 & 181793  & 66.88 \\
(160,160) & 0.130 & 2.58   & 220193  & 59.88 & 0.132 & 1.08   & 246128 & 55.16 & 0.136 & -1.66 & 220193  & 59.88 \\
(128,192) & 0.130 & 2.97   & 228385  & 58.39 & 0.131 & 1.68   & 254320 & 53.66 & 0.134 & -0.40 & 228385  & 58.39 \\
(192,128) & 0.135 & -0.74  & 228385  & 58.39 & 0.135 & -1.33  & 254320 & 53.66 & 0.139 & -3.80 & 228385  & 58.39 \\
(192,192) & 0.133 & 0.64   & 313377  & 42.91 & 0.134 & -0.18  & 350648 & 36.11 & 0.136 & -1.65 & 313377  & 42.91 \\
(128,256) & 0.128 & 3.91   & 346145  & 36.94 & 0.132 & 1.24   & 383416 & 30.14 & 0.135 & 0.90  & 346145  & 36.94 \\
(256,128) & 0.133 & 0.35   & 346145  & 36.94 & 0.137 & -2.56  & 383416 & 30.14 & 0.141 & -5.67 & 346145  & 36.94 \\
\bottomrule
\end{tabular}%
}
\end{table*}

For \ac{SE} after the \ac{BiLSTM} (pre-\ac{FC}), larger hidden-size configurations, such as $(128,160)$, are needed to achieve competitive accuracy ($\mathrm{RMSE}=0.130$), reflecting the reduced influence of post-recurrent recalibration on the temporal features. Applying \ac{SE} after the \ac{FC} layer is largely insensitive to hidden-size scaling, as even the best configuration, $(64,128)$, yields only marginal improvement ($\mathrm{RMSE}=0.133$), indicating that recalibration at this stage cannot compensate for errors accumulated during sequence encoding.

Overall, the analysis shows that asymmetric hidden-unit allocation before the \ac{BiLSTM} maximizes predictive performance while maintaining parameter efficiency, the $(64,128)$ configuration emerges as the optimal design, offering the best balance between \ac{RMSE} reduction and computational cost, making it the preferred choice for edge-deployable channel-gain prediction in resource-constrained scenarios.

\begin{table}[t]
\centering
\caption{End-to-end prediction performance under different training strategies vs. baseline BiLSTM and SE-BiLSTM.}
\label{tab:e2e_performance}
\begin{tabular}{|l|c|c|c|}
\hline
\textbf{Model} & \textbf{Training Strategy} & \textbf{RMSE} & \textbf{$\Delta$ RMSE(\%) } \\ \hline
BiLSTM & \makecell{Without AE \\ (Baseline)} & 0.134 & 0.00 \\ \hline
\rowcolor{green!20} SE-BiLSTM & Without AE & \textbf{0.127} & \textbf{+5.03} \\ \hline
AE $\rightarrow$ BiLSTM & Independent & 0.152 & -13.36 \\ \hline
AE $\rightarrow$ SE-BiLSTM & Independent & 0.146 & -9.07 \\ \hline
\rowcolor{green!20} AE $\rightarrow$ SE-BiLSTM & Compression-aware & \textbf{0.142} & \textbf{-6.58} \\ \hline
AE $\rightarrow$ SE-BiLSTM & End-to-end & 0.166 & -24.02 \\ \hline
\end{tabular}
\end{table}

\subsection{End-to-end prediction performance}\label{sec:res:e2e}
We evaluate the \ac{E2E} performance of the full \ac{LITE} pipeline, where the \ac{AE} encoder--decoder and the \ac{SE}-enhanced \ac{BiLSTM} predictor operate jointly under a fixed 50\% \ac{CSI} compression ratio. Table~\ref{tab:e2e_performance} summarizes the impact of different training strategies described in Section~\ref{subsec:e2e} (independent, compression-aware, and fully end-to-end), compared against the \ac{BiLSTM} baseline and the lightweight \ac{SE}-\ac{BiLSTM} trained on the original dataset (i.e., without \ac{AE} compression/decompression) as references.

Introducing the \ac{AE} inevitably degrades performance due to reconstruction artifacts. In the independent-training setting, where the \ac{AE} and the predictor are trained separately using uncompressed \ac{CSI}, the \ac{AE}+\ac{BiLSTM} and \ac{AE}+\ac{SE}-\ac{BiLSTM} configurations yield \ac{RMSE} values of 0.152 (\textbf{-13.36\%}) and 0.146 (\textbf{-9.07\%}), respectively. These results highlight the mismatch between separate training and joint inference conditions.

\begin{table}[t]
\centering
\caption{GPU memory usage of the TensorRT‑optimized BiLSTM and SE‑BiLSTM engines during deployment.}
\label{tab:trt_memory_usage}
\begin{tabular}{|l|c|c|}
\hline
\textbf{Metric} & \textbf{BiLSTM} & \textbf{SE-BiLSTM} \\ \hline
Engine Device Memory [MiB] & 12.06 & 24.06 \\ \hline
\end{tabular}
\end{table}

\begin{table}[t]
\caption{Latency–throughput trade-offs for TensorRT-optimized BiLSTM and SE-BiLSTM. Latency per sample is measured with batch size 250; improvement (\%) is relative to the non-optimized BiLSTM baseline.}
\label{tab:TR}
\centering
\resizebox{\columnwidth}{!}{%
\begin{tabular}{|l|c|c|c|c|}
\hline
\textbf{Model} & \textbf{Optimized} & \textbf{Lat/sample (ms)} & \textbf{QPS} & \textbf{Imp. (\%)} \\ \hline
BiLSTM         & No                 & 0.03155                 & 31697        & 0       \\ \hline
BiLSTM         & Yes                & 0.01775                 & 56332        & +77.7   \\ \hline
SE-BiLSTM      & No                 & 0.01174                 & 85157        & +168.6  \\ \hline
\rowcolor{green!20} SE-BiLSTM      & Yes                & 0.00679                 & 147267       & +364.5  \\ \hline
\end{tabular}%
}
\end{table}

The compression-aware strategy addresses this issue by training the SE-\ac{BiLSTM} on \ac{AE}-decoded trajectories, allowing it to adapt to distortions introduced during reconstruction. This improves performance to \textbf{RMSE = 0.142} (\textbf{-6.58\%}), reducing the accuracy degradation and confirming that separating reconstruction adaptation from temporal prediction provides the most robust outcome under fixed-compression constraints. 

Fully end-to-end training underperforms, producing \textbf{RMSE = 0.166} (\textbf{-24.02\%}), due to unstable gradients and conflicting objectives between reconstruction and forecasting. These results corroborate that, for aggressive compression, disentangling the learning of reconstruction and temporal prediction is more effective than joint optimization.

\subsection{Memory and prediction time at deployment}\label{sec:res:latency}
While previous Sections focuses on lowering model memory requirements at inference through the use of an efficient attention mechanism (Section \ref{sec:res:sebilstm} and \ref{sec:res:e2e}), this section analyzes the impact of model architecture and optimization on memory footprint and inference latency at deployment.

We evaluated the \ac{BiLSTM} and \ac{SE}-\ac{BiLSTM} models using identical TensorRT configurations (trtexec) with FP16 mixed precision and matched dynamic shape profiles. Table~\ref{tab:trt_memory_usage} reports the persistent device memory allocated by the TensorRT engines during runtime.

The \ac{SE}-\ac{BiLSTM} engine requires 24.06~MiB of device memory, compared to 12.06~MiB for the BiLSTM, representing an increase of approximately $2\times$. In TensorRT, device memory at runtime includes not only the model weights, but also persistent state and enqueue memory for intermediate activations and scratch buffers required during network execution. 


Focusing on latency and throughput, Table~\ref{tab:TR} summarizes the results for both models under TensorRT-optimized inference with a fixed batch size of 250 sequences. For the TensorRT-optimized \ac{BiLSTM} model, the latency per sample decreases to 0.01775~ms, corresponding to a throughput of 56332~\ac{QPS} and a 77.7\% improvement relative to the non-optimized baseline. The inference throughput, measured in \ac{QPS}, is computed as $1/ls$, where $ls$ is latency per sample (s).

The TensorRT-optimized SE-BiLSTM achieves a latency per sample of 0.00679~ms and a throughput of $147267$~\ac{QPS}, which corresponds to a $364.5\%$ improvement relative to the BiLSTM non-optimized baseline. Compared to the non-optimized SE-BiLSTM throughput of 85157~\ac{QPS}, TensorRT provides an additional throughput gain of $72.9\%$.

\begin{table*}[t]
\centering
\caption{End-to-end effect of dataset size \(N\) on trajectory diversity (Pearson correlation), AE reconstruction, and predictor performance. For each \(N\), the same AE feeds all predictors. Highlighted row (\(N=2500\)) shows the best trade-off between reconstruction accuracy and effective diversity under 50\% \ac{CSI} compression.}
\label{tab:end_to_end_results}
\resizebox{\textwidth}{!}{%
\begin{tabular}{|c|ccc|ccc|cccccc|}
\hline
\multirow{3}{*}{\textbf{N}} 
& \multicolumn{3}{c|}{\textbf{Pearson Corr.}} 
& \multicolumn{3}{c|}{\textbf{Autoencoder}} 
& \multicolumn{6}{c|}{\textbf{Prediction RMSE (Same AE per N)}} \\ 

& \textbf{Mean} & \textbf{Std} & \textbf{Median}
& \textbf{MSE} & \textbf{RMSE} & \textbf{\(R^2\)}
& \multicolumn{2}{c}{\textbf{Baseline (without AE)}}
& \multicolumn{2}{c}{\textbf{Independent}}
& \textbf{Comp.-aware}
& \textbf{End-to-end} \\

&  &  &  
&  &  &  
& \makecell[c]{\textbf{SE-BiLSTM}}
& \makecell[c]{\textbf{BiLSTM}}
& \makecell[c]{\textbf{AE$\rightarrow$BiLSTM}}
& \makecell[c]{\textbf{AE$\rightarrow$SE-BiLSTM}}
& \makecell[c]{\textbf{AE$\rightarrow$SE-BiLSTM}}
& \makecell[c]{\textbf{AE$\rightarrow$SE-BiLSTM}} \\ 
\hline

200  
& 0.513 & 0.079 & 0.529 
& 0.352 & 0.593 & 0.636
& 0.213 & 0.220 & 0.575 & 0.570 & 0.555 & 0.567 \\ \hline

1000 
& 0.689 & 0.158 & 0.733 
& 0.035 & 0.187 & 0.964
& 0.131 & 0.143 & 0.216 & 0.211 & 0.208 & 0.212 \\ \hline

1500 
& 0.714 & 0.137 & 0.746 
& 0.025 & 0.158 & 0.975
& 0.143 & 0.154 & 0.215 & 0.210 & 0.210 & 0.205 \\ \hline

2000 
& 0.712 & 0.133 & 0.745 
& 0.018 & 0.133 & 0.982
& 0.136 & 0.142 & 0.180 & 0.179 & 0.174 & 0.190 \\ \hline

\rowcolor{green!20}
\textbf{2500} 
& \textbf{0.707} & \textbf{0.132} & \textbf{0.738}
& \textbf{0.009} & \textbf{0.094} & \textbf{0.991}
& \textbf{0.127} & \textbf{0.134}
& \textbf{0.152} & \textbf{0.146}
& \textbf{0.142}
& \textbf{0.166} \\ \hline

3000 
& 0.787 & 0.056 & 0.794 
& 0.002 & 0.047 & 0.998
& 0.181 & 0.183 & 0.189 & 0.187 & 0.183 & 0.218 \\ \hline

3500 
& 0.776 & 0.060 & 0.785 
& 0.001 & 0.037 & 0.999
& 0.182 & 0.182 & 0.186 & 0.186 & 0.181 & 0.208 \\ \hline

4000 
& 0.793 & 0.053 & 0.800 
& 0.001 & 0.028 & 0.999
& 0.188 & 0.192 & 0.194 & 0.190 & 0.188 & 0.220 \\ \hline

\end{tabular}}
\end{table*}

\begin{figure}[t]
 \centering
\includegraphics[width=0.5\textwidth]{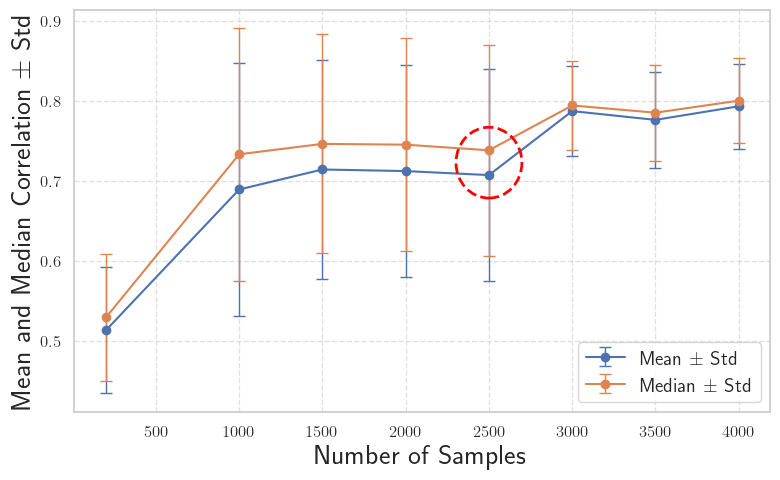}
\caption{Step‑wise Pearson correlation of CSI trajectories for increasing dataset sizes N.}
 \label{fig:Person}
\end{figure}

Although the SE-BiLSTM exhibits an approximately $2\times$ higher TensorRT engine memory footprint, it remains within the capacity of typical Near-RT RIC platforms. While this increase may affect multi-model deployment on resource-constrained systems, the efficient squeeze-and-excitation mechanism and TensorRT optimization enable substantially lower inference latency and higher throughput. These results indicate that the LITE architecture is well-suited for edge deployment scenarios requiring low-latency, high-throughput \ac{CSI} prediction.

\subsection{Data Augmentation: Impact on AE and Predictor Performance}\label{sec:res:csi-compre}

The \ac{AE} plays a central role in this framework by compressing the \ac{CSI}, reducing the communication overhead over the \ac{X-haul} while enabling accurate reconstruction at the receiver. To achieve this, it must be trained on a sufficiently large and diverse set of trajectories. Excessively correlated synthetic trajectories, however, induce averaging behavior during training, which degrades the reconstruction of fine-grained channel-gain variations and limits the usefulness of the compressed representation for both accurate recovery and efficient transmission.

Table~\ref{tab:end_to_end_results} and Fig.~\ref{fig:Person} summarize the impact of dataset size \(N\) on trajectory correlation (measured using Pearson correlation), AE reconstruction performance, and downstream prediction error. For small datasets, such as \(N=200\), corresponding to the largest dataset used in \cite{Alonso2024}, the low mean correlation (0.513) reflects high variability, but insufficient coverage of the trajectory space results in poor AE reconstruction (MSE 0.352, RMSE 0.593) and lowest performance among all predictors.  

As \(N\) increases to 1000--2000, the mean Pearson correlation rises (0.689--0.712) while sufficient variability remains, leading to substantial improvements in AE reconstruction (RMSE 0.187--0.133) and better predictive performance across the BiLSTM variants. The AE accurately reconstructs the generated sequences, providing high-quality inputs to the predictors while retaining meaningful temporal dynamics.

The best trade-off between dataset size and effective trajectory diversity is observed at \(N=2500\). Here, the AE achieves very low reconstruction error (RMSE 0.094, \(R^2 = 0.991\)) and the predictors reach optimal performance (BiLSTM RMSE 0.134, SE-BiLSTM RMSE 0.127) while the mean Pearson correlation remains moderate (0.707) with non-negligible variance. This indicates that the dataset is sufficiently large for accurate learning yet still preserves trajectory variability, allowing the BiLSTM models to capture relevant temporal patterns effectively.

For \(N > 2500\), AE reconstruction continues to improve (RMSE 0.047--0.028) and mean correlation increases (0.776--0.793), but the variance of the correlation drops sharply, indicating highly homogeneous trajectories. This homogenization reduces the effective diversity of the inputs: sequences are reconstructed accurately, yet their temporal nuances become too uniform for the BiLSTM to extract additional information, leading to plateaued or slightly degraded predictor performance (e.g., SE-BiLSTM RMSE 0.181--0.188 for \(N=3000\)–4000).  

These observations align with prior studies on synthetic time series and sequence modeling~\cite{chen2024diversitysyntheticdataimpact}, which show that adding synthetic samples without preserving diversity can produce redundant patterns that limit predictors' ability to capture temporal dynamics. In our case, this explains why increasing \(N\) beyond 2500 improves AE reconstruction but does not further benefit BiLSTM performance, supporting the choice of \(N=2500\) as the balanced dataset size for evaluating LITE.

\section{Conclusion and Future Work}\label{sec:conclusion}
This work introduced LITE, a lightweight end-to-end pipeline for trajectory-unaware channel-gain prediction in CF-MaMIMO systems under O-RAN constraints. By combining a compact 1-D convolutional autoencoder at the O-DU with an asymmetric SE-enhanced BiLSTM at the Near-RT RIC, LITE reduces X-haul transport load and computational footprint while maintaining short-horizon prediction accuracy. The evaluation demonstrates that: (\textit{i}) asymmetric SE-BiLSTM architectures improve accuracy with significantly lower model complexity compared to symmetric baselines; (\textit{ii}) compression-aware training effectively compensates for AE-induced distortions, limiting accuracy loss to 6\% under a fixed 50\% CSI compression ratio versus the BiLSTM baseline; and (\textit{iii}) a TensorRT implementation delivers a 4.6$\times$ throughput improvement, enabling real-time inference at the RIC. Collectively, these results show that LITE provides a practical, deployment-aligned solution for mobility-driven channel prediction in open, disaggregated RAN environments.

Future research directions include: (\textit{i}) evaluating LITE on real dynamic CSI traces to assess robustness under realistic propagation and hardware conditions; (\textit{ii}) jointly optimizing compression ratio and predictor architecture for adaptive X-haul utilization based on traffic and mobility; (\textit{iii}) integrating LITE into closed-loop Near-RT RIC control pipelines, e.g., mobility-aware clustering, handover optimization, or beam management, to demonstrate system-level gains; and (\textit{iv}) exploring model quantization, pruning, and hardware-aware NAS to further reduce the SE-BiLSTM memory footprint, enabling execution on resource-constrained RIC platforms or DUs.


\bibliographystyle{IEEEtran}
\bibliography{main.bib}

@IEEEtranBSTCTL{IEEEexample:BSTcontrol,
CTLuse_forced_etal = "yes",
CTLmax_names_forced_etal = "3",
CTLnames_show_etal = "2",
}

@misc{ericsson2025mobilityreport,
  title        = {Ericsson Mobility Report: November 2025},
  author       = {Ericsson},
  year         = {2025},
  month        = {November},
  howpublished = {\url{https://www.ericsson.com/en/reports-and-papers/mobility-report/reports/november-2025}},
  note         = {Accessed 10 January 2026}
}

@ARTICLE{Mohammadi2024,
  author={Mohammadi, Mohammadali and Mobini, Zahra and Quoc Ngo, Hien and Matthaiou, Michail},
  journal={Proceedings of the IEEE}, 
  title={Next-Generation Multiple Access With Cell-Free Massive MIMO}, 
  year={2024},
  volume={112},
  number={9},
  pages={1372-1420},
  doi={10.1109/JPROC.2024.3451372}}

@ARTICLE{Björnson2020,
  author={Björnson, Emil and Sanguinetti, Luca},
  journal={IEEE Transactions on Communications}, 
  title={Scalable Cell-Free Massive MIMO Systems}, 
  year={2020},
  volume={68},
  number={7},
  pages={4247-4261},
  doi={10.1109/TCOMM.2020.2987311}}

@article{Chen2022,
title = {A survey on user-centric cell-free massive MIMO systems},
journal = {Digital Communications and Networks},
volume = {8},
number = {5},
pages = {695-719},
year = {2022},
issn = {2352-8648},
doi = {https://doi.org/10.1016/j.dcan.2021.12.005},
url = {},
author = {Shuaifei Chen and Jiayi Zhang and Jing Zhang and Emil Björnson and Bo Ai}
}

@ARTICLE{Chen2021,
  author={Chen, Shuaifei and Zhang, Jiayi and Björnson, Emil and Zhang, Jing and Ai, Bo},
  journal={IEEE Journal on Selected Areas in Communications}, 
  title={Structured Massive Access for Scalable Cell-Free Massive MIMO Systems}, 
  year={2021},
  volume={39},
  number={4},
  pages={1086-1100},
  doi={10.1109/JSAC.2020.3018836}}

@ARTICLE{Jiang2022,
  author={Jiang, Hao and Cui, Mingyao and Ng, Derrick Wing Kwan and Dai, Linglong},
  journal={IEEE Journal on Selected Areas in Communications}, 
  title={Accurate Channel Prediction Based on Transformer: Making Mobility Negligible}, 
  year={2022},
  volume={40},
  number={9},
  pages={2717-2732},
  doi={10.1109/JSAC.2022.3191334}}

@ARTICLE{Mismar2024,
  author={Mismar, Faris B. and Kaya, Aliye Ozge},
  journal={IEEE Networking Letters}, 
  title={Adaptive Compression of Massive MIMO Channel State Information With Deep Learning}, 
  year={2024},
  volume={6},
  number={4},
  pages={267-271},
  doi={10.1109/LNET.2024.3475269}}

@misc{DeBast2021,
doi = {10.21227/nr6k-8r78},
url = {https://dx.doi.org/10.21227/nr6k-8r78},
author = {Sibren De Bast and Sofie Pollin},
publisher = {IEEE Dataport},
title = {Ultra Dense Indoor MaMIMO CSI Dataset},
year = {2021} }

@INPROCEEDINGS{Camelo2019,
  author={Camelo, Miguel and Shahid, Adnan and Fontaine, Jaron and de Figueiredo, Felipe Augusto Pereira and De Poorter, Eli and Moerman, Ingrid and Latre, Steven},
  booktitle={2019 IEEE International Symposium on Dynamic Spectrum Access Networks (DySPAN)}, 
  title={A semi-supervised learning approach towards automatic wireless technology recognition}, 
  year={2019},
  volume={},
  number={},
  pages={1-10},
  doi={10.1109/DySPAN.2019.8935690}}

@ARTICLE{Fontaine2020,
  author={Fontaine, Jaron and Ridolfi, Matteo and Van Herbruggen, Ben and Shahid, Adnan and De Poorter, Eli},
  journal={IEEE Access}, 
  title={Edge Inference for UWB Ranging Error Correction Using Autoencoders}, 
  year={2020},
  volume={8},
  number={},
  pages={139143-139155},
  doi={10.1109/ACCESS.2020.3012822}}

@ARTICLE{Beerten2025,
  author={Beerten, Robbert and Ranjbar, Vida and Guevara, K Andrea P. and Pollin, Sofie},
  journal={IEEE Open Journal of the Communications Society}, 
  title={Mobile Cell-Free Massive MIMO: A Practical O-RAN-Based Approach}, 
  year={2025},
  volume={6},
  number={},
  pages={593-610},
  doi={10.1109/OJCOMS.2024.3523217}}

@INPROCEEDINGS{Girycki2024,
  author={Girycki, Adam and Rahman, Md Arifur and Guevara, Andrea P. and Pollin, Sofie},
  booktitle={2024 Joint European Conference on Networks and Communications \& 6G Summit (EuCNC/6G Summit)}, 
  title={Beamforming and Functional Split Selection for Scalable Cell-free mMIMO Networks}, 
  year={2024},
  volume={},
  number={},
  pages={1-6},
  doi={10.1109/EuCNC/6GSummit60053.2024.10597132}}

@ARTICLE{Liu2020,
  author={Liu, Pei and Luo, Kai and Chen, Da and Jiang, Tao},
  journal={IEEE Transactions on Wireless Communications}, 
  title={Spectral Efficiency Analysis of Cell-Free Massive MIMO Systems With Zero-Forcing Detector}, 
  year={2020},
  volume={19},
  number={2},
  pages={795-807},
  doi={10.1109/TWC.2019.2948841}}

@ARTICLE{Alonso2024,
  author={Martinez Alonso, Rodney and Beerten, Robbert and Colpaert, Achiel and Guevara, Andrea P. and Pollin, Sofie},
  journal={IEEE Open Journal of the Communications Society}, 
  title={Trajectory-Unaware Channel Gain Forecast in a Distributed Massive MIMO System Based on a Multivariate BiLSTM Model}, 
  year={2024},
  volume={5},
  number={},
  pages={5348-5363},
  doi={10.1109/OJCOMS.2024.3451313}}

@INPROCEEDINGS{Liu2022,
  author={Liu, Shicong and Gao, Zhen and Hu, Chun and Tan, Shufeng and Fang, Liang and Qiao, Li},
  booktitle={2022 International Wireless Communications and Mobile Computing (IWCMC)}, 
  title={Model-Driven Deep Learning Based Precoding for FDD Cell-Free Massive MIMO with Imperfect CSI}, 
  year={2022},
  volume={},
  number={},
  pages={696-701},
  doi={10.1109/IWCMC55113.2022.9825064}}

@INPROCEEDINGS{Hu2018,
  author={Hu, Jie and Shen, Li and Sun, Gang},
  booktitle={2018 IEEE/CVF Conference on Computer Vision and Pattern Recognition}, 
  title={Squeeze-and-Excitation Networks}, 
  year={2018},
  volume={},
  number={},
  pages={7132-7141},
  doi={10.1109/CVPR.2018.00745}}

@ARTICLE{Schuster1997,
  author={Schuster, M. and Paliwal, K.K.},
  journal={IEEE Transactions on Signal Processing}, 
  title={Bidirectional recurrent neural networks}, 
  year={1997},
  volume={45},
  number={11},
  pages={2673-2681},
  doi={10.1109/78.650093}}

@article{camelo2020ai,
  title={An ai-based incumbent protection system for collaborative intelligent radio networks},
  author={Camelo, Miguel and Mennes, Ruben and Shahid, Adnan and Struye, Jakob and Donato, Carlos and Jabandzic, Irfan and Giannoulis, Spilios and Mahfoudhi, Farouk and Maddala, Prasanthi and Seskar, Ivan and others},
  journal={IEEE Wireless Communications},
  volume={27},
  number={5},
  pages={16--23},
  year={2020},
  publisher={IEEE}
}

@misc{chen2024diversitysyntheticdataimpact,
      title={On the Diversity of Synthetic Data and its Impact on Training Large Language Models}, 
      author={Hao Chen and Abdul Waheed and Xiang Li and Yidong Wang and Jindong Wang and Bhiksha Raj and Marah I. Abdin},
      year={2024},
      eprint={2410.15226},
      archivePrefix={arXiv},
      primaryClass={cs.CL},
      url={}, 
}
\vspace{12pt}
\color{red}
\end{document}